# On-Disk Data Processing: Issues and Future Directions

Mayank Mishra & Arun K. Somani

Department of Electrical and Computer Engineering

Iowa State University

maymishr@gmail.com, arun@iastate.edu

**Abstract**

In this paper, we present a survey of "on-disk" data processing (ODDP). ODDP, which is a form of near-data processing, refers to the computing arrangement where the secondary storage drives have the data processing capability. Proposed ODDP schemes vary widely in terms of the data processing capability, target applications, architecture and the kind of storage drive employed. Some ODDP schemes provide only a specific but heavily used operation like sort whereas some provide a full range of operations. Recently, with the advent of Solid State Drives, powerful and extensive ODDP solutions have been proposed. In this paper, we present a thorough review of architectures developed for different on-disk processing approaches along with current and future challenges and also identify the future directions which ODDP can take.

## 1 Introduction

On-disk data processing refers to the arrangement where the secondary storage devices are also equipped with computing and memory capacity to process stored data. On-disk data processing, or ODDP in short, is an active area of research for the last two decades using Hard Disk Drives (HDD) [1], [2], [3]. Recently, researchers have also proposed ODDP capable Solid State Drives (SSD) [4], [5], [6], [7]. The focus has been to enable processing of data closer to their location, i.e., on the storage drive[1] itself to avoid the resource intensive data movement between the storage drive and the compute nodes or CPUs and save the time and energy.

ODDP is part of a broader research effort known as Near-Data Processing or NDP as shown in Figure 1. NDP also includes techniques for "In-Memory processing" [8], [9], [10] where, the main memory, i.e. RAM, is equipped with the data processing capability. More recently "In-Storage Processing" [11][12] has been proposed, where, the storage controllers are enabled to process data accessed from the drives before forwarding the partial results to compute nodes. The focus of this paper is review and understand the on-disk processing mechanisms and techniques, their building blocks, and identify further research issues and opportunities.

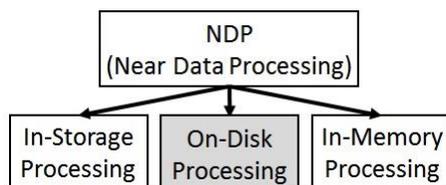

Figure 1: NDP Research Effort Directions.

The purpose of ODDP approaches is to reduce the data movement between the disk and the CPU. The argument is that even a partial filtration/processing of data at a place closer to its location can save a considerable amount of I/O bandwidth, energy, time and other resources. This significantly improves the overall efficiency and performance of the system [2], [3] in more than one aspect.

### 1.1 On-Disk Data processing Vs Normal Data Processing

---

[1] In this document the terms secondary storage, storage drives, storage disks, disk, drives are used interchangeably according to context. They all refer to secondary storage like HDD or SSD.

Figure 2 shows the difference between normal data processing and ODDP. Consider a simple example of a business analytics application which analysis transaction logs of a super market and generates report containing hourly, daily and weekly revenue earned over a given time period. The transaction logs contain time stamped information of the items purchased, their respective prices, discounts offered and final transaction amount in form of a comma separated text file.

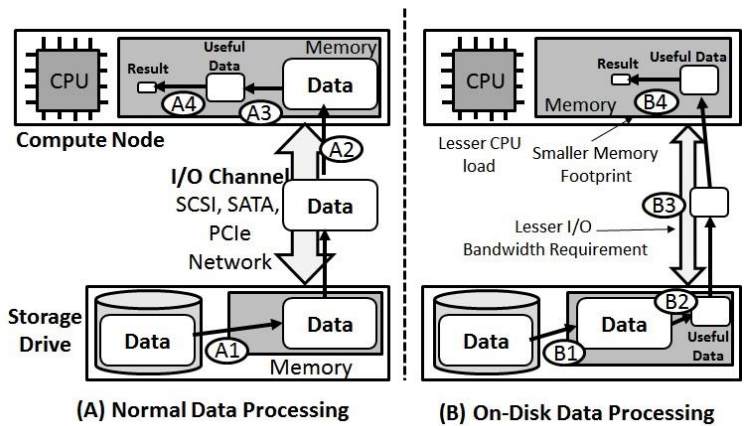

Figure 2: Data Flow in Normal and On-Disk Data Processing.

Figure 2(A) shows the case of normal data processing, where the whole transaction log file from storage drive is accessed and transferred over I/O channels to the RAM of the compute node as shown in steps A1 and A2. The useful data columns containing a time stamp and total transaction amount are then filtered in step A3. Finally, in step A4 the revenues earned are calculated and presented to the user. It can be observed that even though only a part of the overall transaction log is required, the whole file is transferred over the I/O channels and placed in memory. Such an arrangement unnecessarily consumes I/O bandwidth and has high memory footprint.

On the other hand, in ODDP approach shown in Figure 2(B), the storage drives access data in step B1 and then filters out the unnecessary data columns present in the transaction log in step B2. Filtering out unnecessary columns reduces the amount of data to be transferred over I/O channel as shown in step B3. Finally, in step B4 the revenues are calculated. The memory foot print at the compute node is also reduced.

## 1.2 Motivation Behind ODDP

Authors in [1], [2], [13] proposed ODDP model to increase the computing capacity of the system. They argued that the aggregate computing power available in HDDs is more than the aggregate computing power of the compute nodes. To exploit such scenario authors in [1] proposed the offloading of the bulk of data processing to disks and use compute node processor for scheduling, coordination and result aggregation from individual disks. ODDP model was claimed to be better equipped to scale up the processing power with an increase in the size of data. Larger data requires more disks which in turn provide more aggregate processing power.

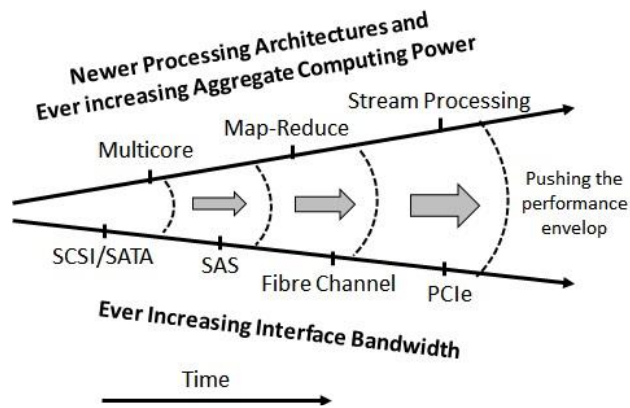

Figure 3: Developments making I/O faster.

However, a working product implementing ODDP model never became available in the market. Newer computing technologies and paradigms continuously increased both the computing capacity and the I/O bandwidth available for data processing and discouraged the effort required to implement an ODDP system.

Figure 3 shows how paradigms like multicore processors, GPUs, and Map-reduce greatly increased the data processing capacity of systems. The newer I/O interfaces coupled with improvement in network bandwidth provided ample capacity to transport the data from the storage drives to the computing nodes. Of course, energy consumption was not paid any attention. Map-Reduce paradigm also brings computing closer to the location of data which seems similar to ODDP. However, there is no reduction in the amount of data read from storage drives.

### 1.2.1 ODDP and Solid State Drives

ODDP is again gaining interest and become a point of discussion in the research community and storage industry with the advent of SSDs. ODDP suits well to modern SSDs which have much higher internal bandwidth than the I/O interfaces [14]. The I/O interfaces, like SATA/SAS which had sufficient capacity to transfer data from HDDs now are becoming the bottleneck as both processors and SSDs are capable of handling data at a much higher rate. Compared to the HDDs, the latest SSDs have 10x data read/write bandwidth [15],[16]. The SATA/SAS I/O interfaces cannot match the read/write bandwidth.

Newer I/O interconnects like PCIe [17] are much faster than the traditional SATA and SAS interconnects, however, it is expected that they too will become bottleneck soon as flash is getting faster and newer technologies like Phase Change Memory(PCM) [18] are getting production ready. ODDP can act as a mitigator to such bandwidth mismatch by reducing the amount of data to be transferred over I/O interconnects.

Another reason behind the renewed interest in ODDP is the modern large data intensive applications like big data analytics. In such applications, the movement of data overshadows the processing of data, thus making a case for ODDP [19]. Moreover, data processing on storage drives will also reduce the memory and cache footprint of data on the compute nodes which may lead to higher data processing capabilities.

ODDP approach is not suited for all data processing applications [20]. The conditions which make an application suitable for on-disk processing are mentioned below.

1. Application's requirement to process stored data. Shifting processing on disks when the stored data is not utilized does not result in any gains.
2. Application's compute and I/O patterns have characteristics such as data dependent logic with high I/O requirement and simple data processing.

It has been shown that applications which have above-mentioned characteristics, such as data-bases, depict a high-performance gain when ODDP is employed [20]. On the other hand, applications involving graph traversals as part of their data processing framework do not show significant improvement with ODDP. This is because all data are needed at the compute nodes any.

The paper is organized as follows. In Section 2, we present the background of data storage on HDD and SSDs including their architectures and the caparison of I/O speeds and throughputs. Section 3 presents the generic on-disk processing model and requirements from such a data processing scheme. We also describe the steps involved in data processing with on-disk processing model in Section 3. Various methods and techniques developed for implementation of these steps in literature are described in Sections 4, 5, 6, and 7, respectively. In Section 8, 9 and 10, we discuss the architecture developed for different on-disk processing approaches in the literature. We make some concluding remarks in Section 11 and identify the future directions in Section 12.

## 2 Background of Data Storage Drives

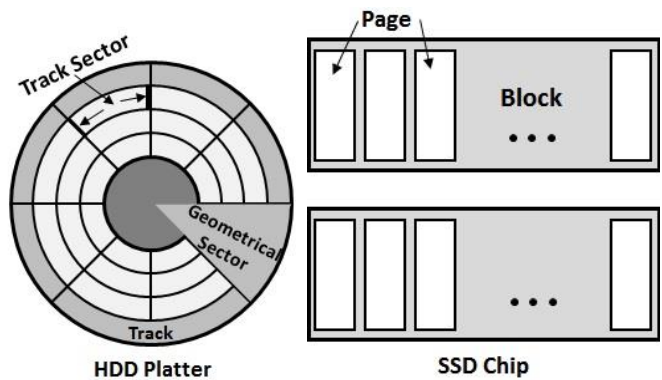

Figure 4: Data Storage arrangement in HDD and SSD.

Before starting the discussion of ODDP, it is necessary to understand the internal architecture of storage drives and how data is assessed over them.

Applications access data in the form of "files", whereas, the data is actually stored on HDDs and SSDs as chunks of 512-4096 bytes each. A file consists of one or more of such data chunks. In HDDs, these data chunks are known as "Track Sectors" or "Sectors" in short and in SSDs the data chunks are known as "Pages" as shown in Figure 4. A sector or a page is the smallest addressable unit of data on disks and is also known as "block" and hence these storages are known as "block devices".

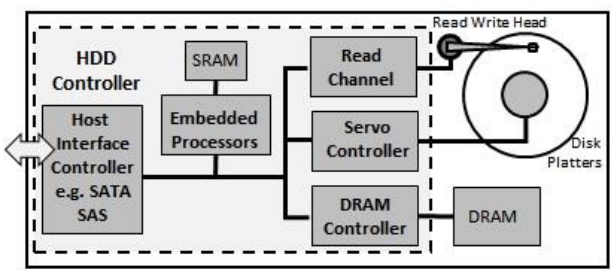

Figure 5: HDD Controller Architecture.

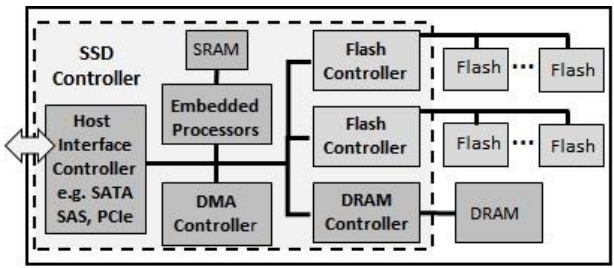

Figure 6: SSD Controller Architecture.

There is both software and hardware infrastructure to support the data access from the storage drives. Figures 5 and 6 show the internal components of HDD and SSD controllers, respectively. The storage drives are tuned to maximize the data I/O performance and were never designed for general purpose computing. A lot of functionalities are thus implemented as firmware and executed on special purpose processors. The main components of the controller include the following.

1. **Host Interface Controller:** It implements a bus interface protocol which carries commands and data to/from the HDD/SSD. HDDs and SSDs both support SATA (Serial Advanced Technology Attachment), SAS (Serial Attached SCSI). The maximum achievable data bandwidth provided by SATA/SAS is close to 600MBps which is sufficient to interface with HDDs. However, SSDs can support much higher data rates. Modern high-performance SSDs have a newer interface named PCIe. PCIe is a newer and much faster interface and supports 5x the data rate than SATA/SAS in modern SSDs.

2. **Embedded Processors:** Embedded processors control the overall functioning of disks and execute the firmware. In the case of SSDs, the firmware consists of Flash Translation Layer (FTL), which is responsible for mapping logical addresses to actual physical Flash addresses. HDDs also have a similar mapping implemented and the execution of the mapping layer is done by the processors.

   The processors are normally low powered RISC processors, like ARM, with multiple cores. For example, Marvell's latest PCIe based SSD controller has 3 ARM Cortex R5 processors each with a clock speed of 500 MHz [21]. SSDs have more such processors than HDDs due to their higher data bandwidth. In HDDs, the processors also interact with read/write head controller and the servo controller to control the positioning of the read/write head over the desired location and control the rotation of disk platters respectively.

In both HDDs and SSDs, the program code and time-critical data are stored in SRAM. The DRAM has higher capacity than SRAM but also has higher latency. Hence, DRAM is used for Data transfers from Flash and disk platters to host and vice versa. In modern PCIe SSDs, the RAM size is in multiple of GBs.

3. **SSD Flash controllers:** The main function of the SSD flash controller is to transfer data between flash memory and the DRAM. There is one flash controller for each flash channel. Essential functions such as error corrections are also implemented inside the controller. Flash controllers employ channel and chip level interleaving for higher I/O performance.

4. **SSD Flash Array:** SSD flash array is a persistent storage medium employing NAND. Each flash chip consists of multiple blocks and each block contains multiple pages. A page is the smallest addressable unit. Reads and Writes are done in units of pages whereas a block is the smallest erasable unit.

Although the architecture of SSD and HDD controllers look similar, their performance is quite different. Table 1 contains the data read, write throughput and latency performance of HDDs and SSDs. The main reason for such a large difference is that in HDDs there is mechanical movement involved in data access. To access a data sector, the read-write head is required to be positioned over a certain track and the disk platter that is rotating to perform the actual read or write over that sector of the track. This results in access latency which is of the order of milliseconds. Moreover, while accessing data stored in non-sequential locations, such mechanical movement is required to serve every request. This results in lower number of such requests being served per unit time. The rate of serving data requests are known as Input Output Operations Per Second or IOPS for short. SSDs are capable of much higher IOPS than HDDs due to all electronic data storage and access.

|  | HDD[15] | SSD[16] |  |
|---|---|---|---|
| I/O Interface | SATA/SAS | SATA/SAS | NVMe |
| Sequential Read (MB/sec) | 230 | 555 | 6,800 |
| Sequential Write (MB/sec) | 230 | 500 | 6,000 |
| Random Read (4KB) IOPS | ~100-200* | 99,000 | 1,000,000 |
| Random Write (4KB) IOPS | ~100-200* | 86,000 | 180,000 |
| Latency Read (us) | Average 4,160 | Max 400 | 100 |
| Latency Write (us) | Average 4,160 | Max 4000 | 30 |
| Capacity | 10 TB | 1.8 TB | 3.2 TB |

Table 1: HDD and SSD I/O Comparison. *- HDD IOPS source [41].

## 3 ODDP Model

A typical data processing is generally seen as a single task, however, there are effectively two processing steps involved [20]. In the first step, known as the filtration step, the entire data is accessed and inspected by CPU to extract useful data and organized in a fashion to facilitate the next step. In the second step, the useful and organized data derived in the first step is further processed. Thus, a data processing task $T$ can be seen as a combination of two subtasks: a) Main compute task $T_C$ which processes useful data, and b) Filter task $T_F$, which derives useful data as shown in Figure 7. An example of the filtering step is the "map" phase of a Map-reduce task.

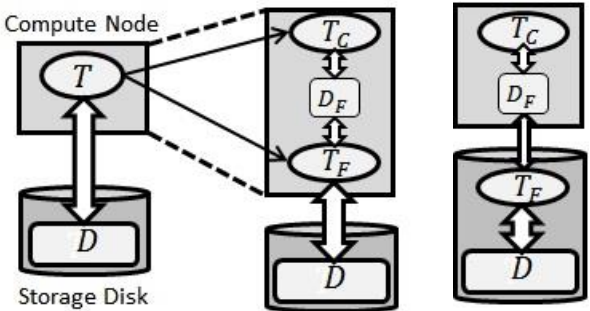

Figure 7: Normal Data Processing.

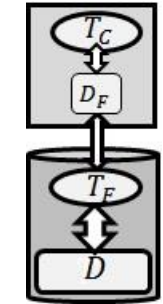

Figure 8: On-Disk Data Processing.

ODDP approaches execute filter task $T_F$ closer to the location of data, i.e., on storage drive itself as shown in Figure 8. In the proposed ODDP approaches filter task $T_F$ can range from complete processing, where, it only sends result to the $T_C$ [4], to partial or pre-processing, where only the useful results extracted from the data are sent to CPU and hence the name "Filtering" or "Shaping" [22].

Unlike complete data processing on the storage drive, "Shaping" and "Filtering" makes data more readily useful and processable for CPU and hence save CPU time. Examples of the shaping/filtering include operations like sorting, encryption/decryption, and selection. In this document, for the ease of description, we will use the term "filtering" to refer to all different levels of data processing done on storage drive by task $T_F$ including the complete processing.

## 3.1 ODDP Challenges

From ODDP user's perspective, offloading filtration task to the storage drive should be seamless and transparent. The user should receive the same level of application execution abstraction from ODDP as received in compute node only execution. ODDP enabled storage drive should seamlessly extend the current application execution model. However, providing such a high-level abstraction is challenging and very wide in scope. A designer and developer of an ODDP solution have to answer following critical questions [1].

a) How are such data processing capable storage drives programmed?

b) What operations are allowed for filtration tasks $T_F$?

c) How the filtration tasks $T_F$ communicate with main tasks $T_C$ executing over compute nodes?

d) How to protect against a buggy filtration task $T_F$ executing over storage drive?

e) What kind of datasets and algorithms can be supported by ODDP model?

The answers to the above questions lie in the nature of target applications. Different applications may require different capabilities from the ODDP enables storage drive. For example, a security application may want to offload operations like encryption, compression, and hashing to storage drives. Whereas BI applications [23], [24] may want storage drives to perform sorting, averaging and other statistical operations.

The key is that if a widely deployed data intensive application requires operations to be performed on ODDP enabled drives, then, ODDP solution can be tailor made for that application. Examples of such ODDP solutions are SSDs performing Sorting [4], Data Merging [25], and Query Processing SSDs [14]**.** The tailor made ODDP solutions are easier to implement when the data processing operations and the structure of data are known.

In certain extreme cases of tailor made ODDP solutions, the filtration operation can even be made available in firmware to speed up the execution. Such solutions do not require any filter programming and deployment by the user and are easier to use. Moreover, providing such solutions is easier for the developers than supporting a general-purpose execution environment on the storage drive. In the next section, we discuss different types of ODDP implementations and the scenarios in which they perform well.

### 3.2 Types of ODDP Implementations

Broadly speaking there are two kinds of ODDP models mentioned below.

1. General Purpose ODDP: Any kind of fitter operations can be performed.
2. 
3. Specific Purpose ODDP: The filter operations available are limited. An extreme case is "single operation ODDP" which performs only a specific and fixed filter operation.

Table 2 below shows the difference between general purpose and specific purpose ODDP models.

|  | **General Purpose** | **Specific Purpose** |
| --- | --- | --- |
| **Examples** | Image Processing | Database Query, Sorting. |
| **Data** | Unstructured, Semi-structured, Structured | Semi-structured or structured data |
| **Suitable For** | Prototyping | Fast Execution |
| **Support for programming** | Yes | No |
| **Filter implementation** | High-level programming library. | Firmware or pre-installed software. |
| **Complexity** | Relatively high | Relatively low |
| **Filter Process Management** | Yes | No |

Table 2: Difference between general purpose and specific purpose ODDP.

We present a survey of various general purpose and specific purpose ODDP approaches developed. We begin the discussion with a review of the requirements for an ODDP solution.

### 3.3 ODDP Requirements

To provide a functional ODDP, a designer is required to provide support for filter task management which involves:

1. **Defining $T_F$:** This step involves describing the data filtering operation. The description can range from a single command like "grep" to a full executable binary depending on the kind of data to be filtered and the application requirements. In general purpose ODDP solutions, the filter is generally defined by the user whereas for specific purpose ODDP solutions such filters generally may be pre-defined and available.

2. **Deploying $T_F$:** Once the filter task is defined, it needs to be placed on disk such that the deployed filter task is invokable by $T_C$. The deployment mechanisms range from dynamic loading of the executable on storage drive in the case of general purpose ODDP solutions to registering with a filter API pre deployed on disk in the case of specific purpose ODDP solution. The data to be filtered needs to be accessible by the filter task.

3. **Invoking $T_F$:** Invocation of a task involves the triggering and execution of the filter task. Filter invocation mechanisms range from explicit triggering by the $T_C$ to automatic triggering when data is available. Section 6 presents different kinds of filter invocation mechanisms employed.

4. **Coordinating $T_F$:** Coordination of filter tasks is required as there can be more than one filter task employed by $T_C$ depending on the nature of data filteration. Moreover, coordination among $T_C$ and the filter tasks is also required because filter tasks are generally not independent tasks. The coordination is in terms of data accesses, data transfers, etc.

In the following four sections, we present the approaches to implement the above-mentioned steps for filter management used by various ODDP schemes in literature.

## 4 Filter Management in ODDP approaches

The filter task $T_F$ can be a standard operation like sort [4], query [14], or can be a complex operation implemented by a full executable binary [20], [22], [26]. Specific purpose ODDP approaches usually employ standard filters and users have a limited or no role in defining them. Users are required to specify or choose a filter among the choices available on storage drive. Such a limited approach is sufficient when the target applications require only a few and standard filters and an assumption about the structure of data can be made, i.e., data is structured or semi structured. In such cases, invokable filters embedded in firmware or a light weight filter library are desirable.

On the other hand, the general purpose ODDM solutions provide a support for the high-level language for users to define the filter tasks. This is because the applications may require a wide variety and changing filter operations executing over unstructured data. In such cases, complex programming constructs may be required to define a filter operation. Thus, the flexibility and power of high-level languages like C, C++, Java or scripting languages like Python, Shell script may be more appropriate.

Another difference between the general purpose and specific purpose ODDP approaches is in the manner filters are deployed. General purpose ODDP approaches prefer to deploy filters dynamically whereas specific purpose ODDP approaches employ mechanisms like pre-installed filter library on the drive or filter code compiled into the drive controller firmware.

A dynamic approach is suited to applications which require different filtering operations on data at different times. Moreover, in the case of multi-application environment where the application remains unknown until the time of actual execution. The pre-installed or firmware compiled filter provides a low overhead but limited filtering solution.

### 4.1 Filter Management In General Purpose ODDP solutions

When high-level languages are supported for defining the filter tasks, there is no restriction of the complexity and capability of the filter to be defined. Such filters provide flexibility to the users. In the following, we present an example of the type of high-level support for filter definition by using examples of the schemes developed in the literature.

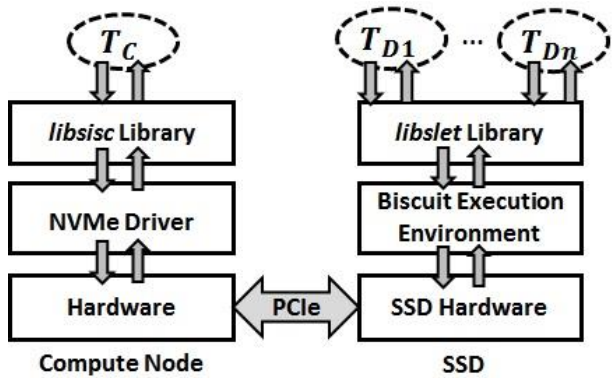

Figure 9: Biscuit's Library Support for Filters.

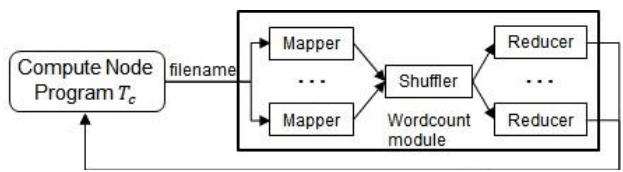

Figure 10: Biscuit Filter program "WordCount".

**4.1.1 Biscuit** [20]**:** Biscuit provides high-level APIs in form of two libraries, namely *libsisc* and *libslet,* which can be employed by programmers. The *libsisc* provides an interface to the tasks executing on compute node and *libslet* provides an interface to the filter tasks executing on SSD. Figure 9 depicts the architecture of the framework. The language supported is C++ and there is no restriction on the kind of data filtering performed.

The filters are defined by inheriting and extending *SSDLet* class. The *SSDLet* class is present both in *libsisc* and *libslet.* However, the one present in libslet library actually defines the working of the filter task $T_F$. The one present in libsisc implements and interface to let compute node task $T_C$ to interact with filter task $T_F$.

As an example, Figure 10 shows the required data processing flow of a map-reduce based word count program implemented using the ODDP functionality presented by Biscuit. All the mapper, shuffler, and reducer tasks are executing over the SSD itself. The task executing on the compute node is only coordinating the data processing tasks on SSD. The code snippets in the table below shows the definition and implementation of the filter and compute node task.

**Filter Definition:** The code snippets shown in Table 3 cells (A), (B), and (C) show the filter definition. In the current example, there are three filter tasks: mapper, shuffler, and reducer which collectively implement the required word count task. The mapper task shown in (A) gets a filename as input from the compute node task. The file contains text and the filter tasks need to count the occurrence of words in the given file. The mapper tokenizes the text and passes tuples <word, 1> to the shuffler.

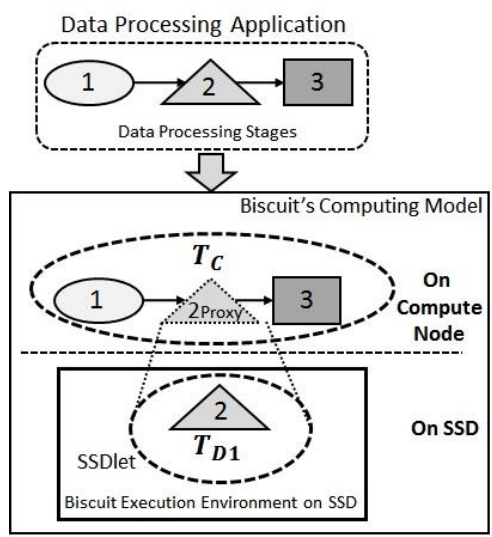

Figure 10: Biscuit's Filter coordination Mechanism.

The communication between mapper and shuffler happens over output port named "output" defined in line 3. These shared ports are actually FIFO producer-consumer queues. Shuffler and Reducers also have input and output ports defined where the output of reducer is sent to the main task executing over compute node.

Biscuit provides a way for these filter tasks to be defined and be contained in a ".slet" file which denotes the SSDlet module. Line 10 of the mapper definition denotes the registering to an SSDLet module. The API for registering is present in *libslet* class.

**Filter Deployment:** The SSDlet module, which is a ".slet" file is placed on the SSD, is loaded using the "loadModule" function call by the main compute node task $T_C$ as shown in line 2 of cell (D). Line 3 denotes the creation of the overall "word count" filter task instance "wc". Lines 4 to 8 denote the creation of instances of individual mapper, shuffler and reducer tasks of "wc".

| $T_F$                                                                                                                                                                                                                                                                                                                                                                                                                                                                                                                                                                         | $T_C$                                                                                                                                                                                                                                                                                                                                                                                                  |
|---|---|
| **Filter Tasks executing over SSD**                                                                                                                                                                                                                                                                                                                                                                                                                                                                                                                                           | **Task executing over compute node**                                                                                                                                                                                                                                                                                                                                                                   |
| (A) **class** Mapper : **public** SSDLet <OUT_TYPE<std::pair<std::string, unit_32_t>>, ARG_TYPE<File>> { **public:**    **void** run() { 1.   **auto&** file = getArgument<0>(); 2.   FileStream fs(std::move(file)); 3.   **auto** output = getOutputPort<0>(); 4.   **while** (**true**) { 5.   … 6.    **if**(!readline(fs, line)) **break**; 7.    line.tokenize(); 8.    **while** ((work = line.next_token()) != line.cend()){     // put output(i.e., each word) to the output port 9.    **if**(!output.put({std::string(word), 1}))       **return**; }}}} // register class in its container module 10.  RegisterSSDLet(idMapper, Mapper) | (D) int main(int argc, char *argv[]) { 11.  SSD ssd("/dev/nvme0n1"); 12.  auto mid = ssd.loadModule(File(ssd, "wordcount.slet")); 13.  //create an Application instance and proxy SSDLet instances 14.  Application wc(ssd); 15.  // Defining Mappers, shufflers & Recucers. 16.  SSDLet mapper1(wc, mid, "idMapper",       make_tuple(File(ssd, filename))); 17.  SSDLet mapper2(…); 18.  … 19.  SSDLet shuffler(wc, mid, "idShuffler"); 20.  SSDLet reducer1(wc, mid, "idReducer"); 21.  SSDLet reducer2(wc, mid, "idReducer"); 22.  … 23.  // Making data passing connections 24.  wc.connect(mapper1.out(0), shuffler.in(0)); 25.  wc.connect(mapper2.out(0), shuffler.in(0)); 26.  … 27.  wc.connect(shuffler.out(0), reducer1.in(0)); 28.  … |
| (B) **class** Shuffler : **public** SSDLet{   **public:**     **void** run() {       **auto** output0 = getOutputPort<0>();       **auto** output1 = getOutputPort<1>();       …       **auto** input = getInputPort<0>();       // Shuffling Code. }}}} RegisterSSDLet(idShuffler, Shuffler) | 29.  auto port1 = wc.connectTo<pair<string,       uint32_t>>(reducer1.out(0)); 30.  auto port2 = wc.connectTo<pair<string,       uint32_t>>(reducer2.out(0)); 31.  // Strat application to invoke and execute SSDLets 32.  Wc.start(); 33.  Pair<string, unint32_t> value; |
| (C) **class** Reducer : **public** SSDLet{   **public:**     **void** run() {       **auto** output = getOutputPort<0>();       **auto** input = getInputPort<0>();       // Reducer Code. }}}} RegisterSSDLet(idReducer, Reducer) | 34.  While(port1.get(Value)|| port2.get(value) || ..) 35.  cout << value.first << "\t"<< value.second << end; 36.  //wait untill all SSDLets stop 37.  wc.wait(); 38.  ssd.unloadModule(mid); 39.  Return 0; } |

Table 3: Biscuit's filter code snippets.

**Filter Coordination:** The interaction between the individual filter tasks of "wc" in terms of data stream passing is enabled in lines 9 to 11 of the main compute node task. The output ports of certain filter tasks are connected to the input ports of other filter tasks to enable the overall data processing operation.

The data input ports of the main compute node task are also connected to the reducers in lines 12 and 13. Figure 10 shows the coordination mechanism. Data processing application is shown as a three stage program. Out of these three stages, the second stage is "wc" implemented as SSDLet. The figure shows how stage 2 is represented by proxy on the compute side process. In code lines, 4 to 8 define these proxies. Any data input to the proxy of stage 2 is sent to the SSDlet automatically by the Biscuit runtime.

**Filter Invocation:** The filter task is invoked by using the "wc.start()" function call on line 14 and the execution begins on SSD. The data returned by the reducer filter tasks is received in line 16 and printed. The received values can further be processed by the compute node task.

Biscuit employs the data flow based processing arrangement as evident from the example presented above. The underlying communication between the filter tasks and the main compute node tasks is enabled by the *libsisc* and *libslet* libraries.

### 4.1.2 MVSS [22]

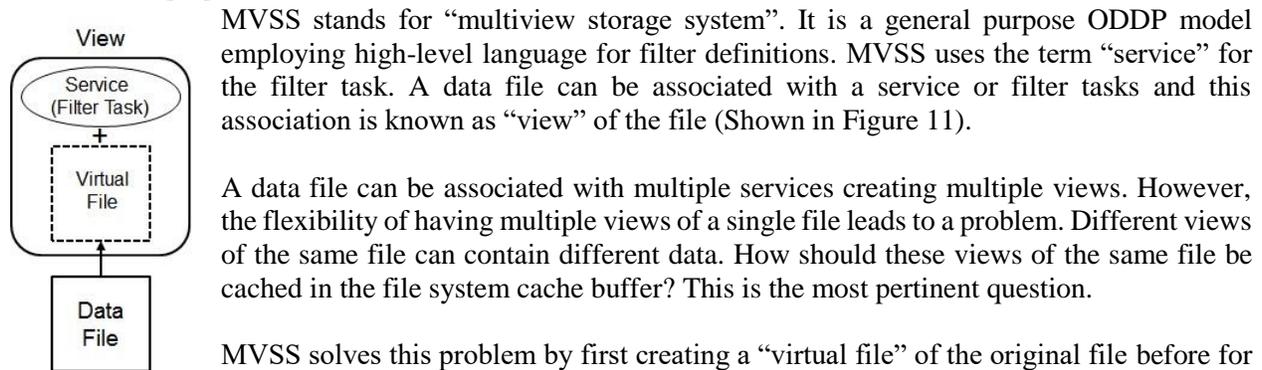

Figure 11: MVSS View.

MVSS stands for "multiview storage system". It is a general purpose ODDP model employing high-level language for filter definitions. MVSS uses the term "service" for the filter task. A data file can be associated with a service or filter tasks and this association is known as "view" of the file (Shown in Figure 11).

A data file can be associated with multiple services creating multiple views. However, the flexibility of having multiple views of a single file leads to a problem. Different views of the same file can contain different data. How should these views of the same file be cached in the file system cache buffer? This is the most pertinent question.

MVSS solves this problem by first creating a "virtual file" of the original file before for every service to be associated. Moreover, every virtual file resides on a separate virtual disk. A virtual disk behaves like a normal data storage device but has no corresponding physical disk. A virtual disk is hooked to the existing physical disk and forwards all I/O requests directed to itself to the actual physical disk. Virtual disks, which are created using "mount" command, provides different namespaces for different virtual files and thus solves the problem of caching. MVSS introduces a stackable file system MVFS (Multiview File System) to manage virtual files.

MVSS makes use of the difference between the actual logical capacity of the storage disk which depends on the address bus width with the real capacity of the disk. Generally, the logical capacity of the disk is many times higher than the actual capacity. MVSS makes use of the logical addresses to locate virtual directories and files.

**Filter Definition:** The services are defined using a high-level language. As MVSS is an ODDP model, it leaves the choice of the language used to the implementations. However, the filter tasks or service follow a data flow mode where they receive input data from one end and produce output data from other. This property of filter tasks in MVSS is similar to Biscuit's filter tasks. An executable implementing the service or filter can be loaded on the disks as any other file. For example, an executable implementing encryption/decryption can be loaded on a disk as a service.

**Filter Deployment:** The filter task or service is deployed by associating the virtual file and the filter task or service using a function call "attach" provided by the MVSS. For example, following function call deploys the filter task *"decrypt"* with decryption key "*dkey*" for a file *"/Vd1/bar"*.

attach (/Vd1/bar, decrypt, dkey)

A service can also be associated with directories using the same "attach" function call. The service gets attached with all the files and subdirectories present in the directory. Different files or directories on a virtual disk can be associated with different services or filter tasks.

**Filter Invocation:** The filter task or service gets invoked whenever the associated virtual file is accessed. The filtered data is returned as a result. Whenever there is an access to the location of virtual file or directory, then, the filter operation is performed on the attached file or directory. The data read from the virtual file or directory results in return of filtered data to the application.

**Filter Coordination:** The data stream property of the filter tasks can also enable the creation of filter pipelines where a filter is applied after other. However, such a pipeline needs to be created by the user by implementing different services in single service or executable. The data stream property of individual filters makes this kind of approach possible.

### 4.1.3 OASIS [26]

OASIS is a proposed ODDP model for object-based storage systems. The object-based storage stores data in form of objects with certain associated meta data or attributes. Unlike hierarchical storage like a file system, the object storage assigns a globally unique identifier to the objects. This identification mechanism provides highly scalable storage and is used for the massive amount of data which is generally unstructured. The metadata stored with the stored data object, called user object, can potentially aid in the filtering. The semantic information stored in metadata can be utilized by the filtering tasks.

Currently, there are 4 different kinds of objects used in OSDs. The data is stored in "user objects" and the addressing and retrieving of user data are performed by using "root object", "partition object" and "collection object". For the purpose of ODDP, OASIS introduces new kind of object called "function object" to offload application functions like compression, classification etc. These function objects can be written in any high-level language like C or JAVA. OASIS proposes a full execution support on Object Storage Device (OSD).

**Filter Definition:** The filters are defined in high any high-level language and even in any scripting language like Python and TCL. Figure 12 shows one example of the filter written in C. The code takes input data in form of stream and also outputs data to a stream. The code filter-out the input values which are greater than or equal to 100 in line 3.

```
    int main(int argc, char *argv[]) {
    Char ch[50];
1.  FILE *instream, *outstream;
    … // open file streams and check error conditions
2.  do{
      fgets(ch, 50, instream)
      // check for NULLs and \n
3.    if(atoi(ch) < 100)
         fprintf(outstream, "%s", ch);
    } while(!feof(instream));
    … //Close streams
    }
```

Figure 12: MVSS filter code snippet.

**Filter Deployment:** Once the filter code is defined it is stored as a function object on the OSD. The OSD needs to have the capability to execute the function object. Moreover, the "user object" which stores the data to be filtered needs to have an entry of the function object ID and parameters (e.g., encryption keys) in its attributes. A filter can be deployed for multiple files by associating the function object to them. Moreover, if a function object is associated with the partition object, the filter will be applied to all the object in that partition. Similarly, all the user objects in an OSD logical unit can be associated with function object by associating function object with root object.

**Filter Invocation:** When the data from the user object is accessed the information in its attributes point to the "function object" storing the filter code. The Function object is then executed over the data content of the "user object" and the results are returned. Such a filtering mechanism is transparent to users.

## 4.2 Filter Management in Specific Purpose ODDP solutions

### 4.2.1 Query Processing on SSDs [14]

This database specific ODDP model proposes offloading of certain selection and aggregation operations involved in database querying to SSD. The proposed model does not support a full set of database operations. The SSD is assumed to contain data in a fixed format in database tables.

**Filter Definition and Deployment:** There are only few filter operations namely selection and aggregation are implemented. The code for these operations is compiled into the firmware of the SSD. Different database operation tasks are registered with different commands. This operation to command registration is part of the code which gets compiled into the firmware. Hence, no new operations or filters can be defined dynamically.

**Filter Invocation:** Filter task consisting of database operations are explicitly invoked by the compute node task by using specifically implemented commands. Another approach is that the filter task gets invoked when data is loaded into RAM from the flash. After the filter operation is performed the data is not explicitly returned to the compute node task. The compute node task uses an explicit "GET" command to retrieve filtered data.

**Filter Coordination:** If multiple filter operations are required then they need to be explicitly invoked by the compute node task. There is an option of implementing multiple operations together as a separate filter task in which case the invocation is required only once.

### 4.2.2 Self-Sorting SSD [4]

Self-Sorting SSD is an extreme case of specific purpose ODDP solution. There is only a single filter operation possible and that is "sort". The nature of data is also assumed to have a key-record pair. The scheme builds an index in SSD which represents the sorted order of data which can be traversed to get sorted data output. Authors also mention that the full sorted order index can also be used to implement functions like range queries, finding min/max and also searching. The index is in the form of B+ trees which results in quick data access and traversal. The index resides in RAM but can also be pushed to the Flash.

**Filter Definition and Deployment:** The sorting algorithm is implemented in the SSD controller firmware itself. There is no option to dynamically change the operation and to load a new filter operation.

**Filter Invocation:** The filter operation, which creates index can be invoked in two different ways. In The first approach, whenever the data is written on the SSD it is parsed and the index is updated. This approach is called "sort-on-write". In the other approach "sort-on-command" the filter operation to create index is explicitly invoked by a specially implemented command. This approach generates an index on demand.

**Filter Coordination:** Since there is only a single filter task which operates across the SSD, there is no need for coordination with any other filter.

### 5 On-disk processing Architecture and Challenges
In the previous section, we discussed the filter management related issues and implementations in literature. The filter management is implemented by the Host Agent which resides on the compute nodes. We discuss the components of both host and disk agents which are involved in enabling ODDP. We use a generic on-disk processing model shown in Figure 13 to present various ODDP components. We first provide a brief explanation of each component and then discuss various approaches to implementing these components in literature.

## 5.1 Host Agent

The host agent provides the compute node side functionality for ODDP. It is usually implemented as a library which the ODDP aware applications utilize. Apart from filter management, as discussed in Section 4, the compute node side task $T_C$ employs host agent for following purposes.

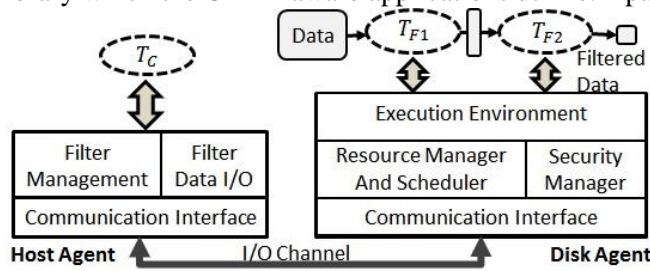

Figure 13: On-Disk Processing Architecture.

1. **Filter Data I/O:** In normal data I/O without ODDP, the data is available to the compute node task after the access request is processed by the storage drives. The data access delay ranges from a few milli-seconds on HDDs to a few hundreds of micro seconds on SSDs (refer Table 1). Moreover, the data received by the compute node tasks is what is asked for. However, in ODDP based data access, a processing delay occurs.

   The processing delay, depending on the size of data, may be much larger than the data access delay in not-ODDP approaches. The size of the data returned by the ODDP enabled storage drive is also generally smaller than the size of actually stored data. Such difference in expected access delay and expected data size may not fit into the semantics of a general data I/O layer on the compute node. The host agent implements functionalities to handle such differences.

2. **Communication Interface:** The host and disk agent together implement the communication interface for ODDP based solutions. The communication interface is utilized for filter management activities like installation and invocation etc. The communication interfaces are also utilized to initiate the data I/O for filtered data, however, it is not used for actual data I/O. Data is still transferred using the traditional data I/O interface as on-ODDP storage drives.

   The communication interface is the most challenging part of any ODDP solution. The challenge is to work with the existing low-level protocols and interface between the applications and the storage drive.

   The current interfaces and protocols only support very basic Data I/O operations like Reads and Writes. Moreover, to optimize the storage drive performance, some of the interfaces and protocols are implemented in firmware and the boundaries between the layers have also been violated. Such mixed implementations and interfaces pose a challenge for any additions or modifications which are necessary to implement ODDP. In Section 6 we discuss how different approaches for implementing the communication interface between the host agent and disk agents have been designed.

## 5.2 Disk Agent

This is the disk side counterpart of the host agent. The disk agents vary greatly in terms of the services and flexibility they provide. The host agents for general purpose ODDP solutions tend to provide much larger functionality than the specific purpose ODDP solutions. Such a difference is natural due to the nature of data processed and kinds of filters supported. The disk agent can also be seen as the operating system of a storage drive. The disk agent provides following main functionalities to support ODDP.

1. **Execution environment:** Ideally the execution environment on storage drive should allow the execution of the filter tasks just like the execution of a normal task on compute nodes. However, due to constrained resources like RAM, small fast memory (SRAM), no memory management unit (MMU), no cache coherence, less than ideal compute power, reduced instruction processors, and restrictive synchronization primitives the execution environment is quite constrained.

Specific purpose ODDP solutions the filter task is actually the extension of the disk management software and may not execute as a separate process. However, in general purpose ODDP solutions, the execution environment requires process semantics, I/O control to let multiple filter tasks execute.

General purpose ODDP solutions allow the execution of filter tasks as separate processes. They provide a set of libraries services which can be invoked by filters defined in high-level languages. For example, Biscuit [20] provides *libslet* library on SSDs which is utilized by filter tasks written in C++. This library acts as a layer between Biscuit's runtime and the filter task. The filter task receives a function table from the runtime which contains pointers to functions or interrupts implementing memory allocation and other I/O related activities. MVSS[22] talks of providing Java Runtime Environment imply.

2. **Resource Manager and Scheduler:** Computing resources on the disks are scares and for efficient utilization the resource management is necessary. Resource manager and scheduler becomes important where multiple filter tasks for more than one applications execute concurrently.

3. **Security Manager:** Due to limited capabilities like locking mechanisms, memory controllers on disks, a malfunctioning filter task can affect the whole disk. To avoid such security threats operations like direct data I/O are not permitted for the filter tasks. They are to be initiated and terminated by the applications executing on compute node. For example in MVSS[22], the I/O requests are passed from the tasks executing over compute nodes to the filter tasks using the IPC (Inter Process Communication) mechanisms.

In Biscuit [20] too filter tasks are prohibited to use low-level block addresses directly and are forced to operate under a file system. Both the libraries provided in biscuit namely *libsisc* and *libslet* implement a file object for this purpose. The file objects are initialized by the main tasks executing on compute node and are passed to the filter tasks. The filter tasks inherit the file access permissions granted to the main task executing on compute node.

In some ODDP approaches, the filter tasks cannot initiate a communication with the main task executing on compute node. Like I/O operations, these are also initiated and terminated by the main tasks executing on compute node.

## 6 Communication Interface

Figure 14 shows the Linux storage stack. There are several software, firmware and hardware layers which together implement the storage stack. Applications interact with storage system in terms of operations on "files". The file system abstracts the block based storages for use by applications. Interface "I1" is the file based interface which supports file operations like "read", "write" etc. The secondary storage device or the block device interact in terms of blocks of data.

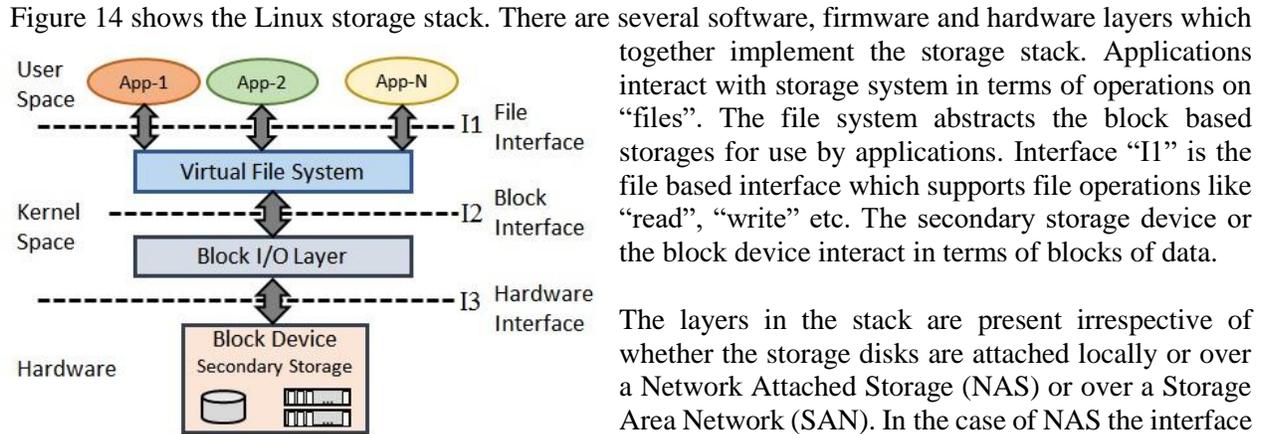

Figure 14: Storage Stack.

The layers in the stack are present irrespective of whether the storage disks are attached locally or over a Network Attached Storage (NAS) or over a Storage Area Network (SAN). In the case of NAS the interface ``I-1" shown in Figure 13 is over a network, and in the case of SAN interface ``I-3" is over the network. Similarly, in object-based storages [27] the interface ``I1" is a key value based and is directly supported by the disks over a network.

The key point to note is that irrespective of the different storage solutions available, the hierarchy of the storage stack layers remains preserved. The difference lies in the location of the layers or functionality, i.e., whether the functionality is implemented on disk or storage controller or over host's operating system.

Applications interact with storage through the Virtual Filesystem (VFS) interface. The Virtual file system is a unified layer which provides a uniform interface of APIs to applications to access various physical files systems like ext4, FAT, NTFS etc. The interface provided by the VFS is limited and only allows applications to perform read and write kind of operations on files. The VFS does not let applications specify the exact data required from the file. For example, consider a text file *file1.txt*, which contains a time-stamped server log, being accessed by an application. Even if the application needs entries for a particular time duration *t1* to *t2*, there is no way for it to mention this during an I/O request.

Another newer storage interface is Object I/O interface [27],[26] where the data is stored in form of objects and accessed using a key. However, even in object-based I/O request, an application is unable to request an element matching a particular value in the object directly.

Apart from the VFS, the block layer interface in the kernel block I/O stack [28], [29] also poses a hurdle in communicating the Data Filtering related information. Currently employed Block I/O interface does not consist of any direct way to specify the filter. This limitation of Block I/O interface is considered as the most critical hurdle in supporting on-disk processing [22], [19], [26]. Currently, there is no standardized approach of passing the filter information along with the I/O request.

These limitations of the VFS and Object I/O interfaces pose a hurdle for ODDP. Applications are not able to specify the location of data to be processed and the kind of filtering or processing to be done over disks. Different solutions to this problem have been proposed in the literature.

**9.1 Using Dedicated Logical Block Addresses (LBA)**

Using dedicated LBAs for communication has been used by many ODDP approaches like MVSS [22] and Self-Sorting SSD [4] which work with the block-based interface of the storage drives. In this approach, some block locations in the storage drives' address space range are reserved for communication purposes. Figure 15 shows the monitored LBA range which is reserved for communication. Whenever the compute node task $T_C$ needs to send a command like invocation of filter task or probing filtered data availability, it writes the message on these reserved LBAs using the normal write operation on a block based storage device.

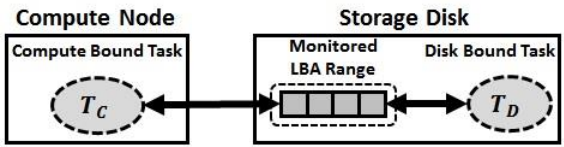

Figure 15: Dedicate LBA Based Communication.

The disk agent monitors these locations and passes the received commands to filter tasks $T_D$. The format of the message passing is not yet standardized and different ODDP approaches employ different messages formats.

The LBA range utilized for the communication may or may not represent the actual physical locations on the disks. This is because the address space range in disks is generally larger than the actual capacities of the disks.

ODDP approaches employing the dedicated LBA range based communication do not expose the LBA range directly to the applications. They provide an interface to the applications and hide the complexity and low-level message passing. The interface provided is generally an extension to the current VFS based data I/O interface.

### 9.1.1 Extending Current File I/O API

MVSS [22] (discussed in Section 4.1.2) proposes a stackable file system [30] which provides an API to the applications to manage filter tasks. For example, it lets the application to associate an operation with a stored file or directory. Figure 16 shows the API in form of ODDP related commands. To support such filter management operations directed by the application MVSS uses back channel communication.

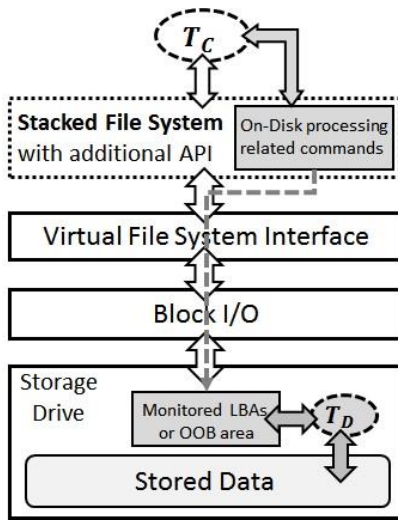

Figure 16: Additional API Based Communication.

The API exposed by the stacked file system (MVFS) makes use IPC (Inter Process Communication) mechanism which is implemented by using dedicated and out of physical range LBAs to send messages to the disk agent. For example, when "attach" API, which associates a filter task with a virtual file, is invoked, the host agent sends a message to disk agent by writing the message on dedicated LBA. The message contains the path of the file and the path of the filter task executable. Similarly, the creation and of virtual file is also performed by passing the messages to the disk agent.

Although the message passing scheme using dedicated LBAs is simple and considered as ``out of band", it is not efficient and may face unnecessary delay due to sharing of the block I/O scheduling. Deciding the LBA range for message passing itself poses challenges due to different disk capacities. Such an arrangement also poses a challenge in the disk to application message passing. For example, if the disk has insufficient memory to carry out the filtration, there is no way for it to inform the application.

### 9.2 Using The Standard I/O Interface

The object-based storage devices (OSD) [31][32][33] hides the block based data I/O interface from the applications. OSDs expose a high-level object-based interface [34][35] which not only consist of basic read and write commands, as available in file systems but also the consists of commands to modify the meta data associated with the user data objects.

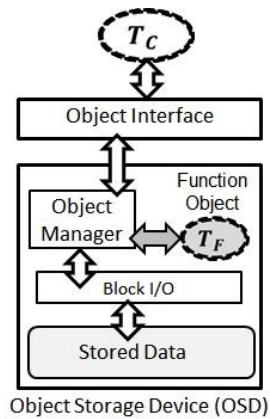

Figure 17: Object I/O Interface.

OASIS [26] (discussed in Section 4.1.3) is an ODDP scheme which utilizes the current object based Interface T10 [35] to manage the associations between a data object and filter task which is implemented as a function object. The filter tasks or the function objects are loaded into the OSD by using the standard object creation commands. Moreover, the association of the filter task with the user data objects is done by including the function object's id to the attribute list or metadata data of user object by using standard attribute modification commands. The function object gets invoked whenever the user object is accessed. The filter tasks present in OSD can also be listed by using standard attribute list command. Thus there is no need of any extra API or a non-standard mechanism for communication between host agent and disk agent for filter management purposes. Figure 17 shows the function object and its interaction with object manager.

Because of their high-level interface and the availability of metadata of user data the Object-based storage devices (OSD) are potentially more capable of on-disk processing than traditional block interface based devices [26].

### 9.3 New Protocol

Authors of Query processing SSD [14] propose a new communication protocol over the SATA/SAS interfaces. The protocol consists of three commands *OPEN, CLOSE* and *GET*. A session starts with an OPEN command and ends with CLOSE. GET command is used to monitor the status of the on-disk processing task and retrieve the results when processing is complete. To enable query processing a proprietary SSD controller firmware is implemented which only understands these three commands.

The corresponding implementation in [14] is only experimental and does not support general query processing. Only a subset of database operations like simple selection and aggregation are implemented as a user program on SSD. The experimental program is triggered by the whenever SSD controller receives commands.

### 9.4 No Communication (Automatic Triggering)

In certain specific purpose ODDP schemes which provide only a single data filtration operation like "sort", there may not be any requirement for extra communication between Td and Tc. For example, self-sorting SSDs proposed in [4], the SSD creates a sorted index on data whenever data is written. Thus, the operation is inbuilt and does not require any explicit communication.

## 10 Scheduling and Resource Management

**Cooperative Multithreading**

Biscuit supports execution of multiple on-disk filtering tasks, called SSDLets, in a concurrent manner. It allocates a schedulable unit called fiber to each SSDLet and the fiber executes the SSDLet code. Biscuit employs cooperative multi threading [36], a limited form of general multithreading, where only a blocking

I/O or an explicit yield call results in switching of context. Such a cooperative approach has low context overhead and no locking is required for resource sharing. This suits well to the resource constrained environment of an SSD.

The current SSDs do not support MMU (Memory Management Unit) or Virtual memory. However, Biscuit locates each filter task instance in a separate address space. Thus multiple instances of same filter task can be supported. Biscuit also supports dynamic memory allocation and maintains a "system memory allocator" which allocates memory for the runtime initiated objects and "user memory allocator" which allocates memory to SSDLet instances. System memory is inaccessible for SSDlets.

**First Come First Serve**

OASIS [26], the object interface based ODDP approach, schedules function objects, i.e. the filter tasks, in first come first serve basis. A function object is scheduled whenever the object storage drive receives a data object access call. Thus, the scheduling of the filter task depends on the object access pattern.

**Time division Scheduling**

Query Processing over SSD [14], which is a specific purpose ODDP solutions executes multiple queries at once by doing a time division multiplexing. There is a single execution thread which cycles over the submitted queries and performs execution for each one of them over specified data in a time slot. If there is a single query then all slots are allocated to that query only.

Such a simple scheduling approach is possible because this ODDP solution implements only selection queries and the code to execute those queries is part of the firmware. The code executing is same for each query, only the data set changes. It can be seen that apart from the time slot there is no other resource allocation required.

**No Scheduling**

Self-Sorting SSD [4] implements only sort operation. Moreover, the operation is performed on whole data stored in SSD. Thus, there is no need for scheduling. The sort operation executes in two modes. One is sort-on-write and another is sort-on-command. As there is only a single command issuing multiple commands does run as a separate instance. There is only one sorting thread at any point in time.

## 11 Challenges to ODDP

### Distributed Data

ODDP offloads data processing operations on storage drives. The execution of such offloaded operations known as filter tasks faces a challenge when the data to be filtered spans across multiple disks and the filter operation needs access to the full data. For example, operating on an encrypted file stored across several storage drives. However, if the distributed data parts are independent of each other the filters can operate on the individual part and the results can later be combined by the compute node task. An example is the Map-Reduce kind of data processing scenario.

### RAID

The RAID controller does a logical remapping (block level or byte level) of data across several storage drives. This mapping is transparent to the application accessing data as the data access request is sent to RAID controller and not directly to the storage drives. The logical remapping of data poses a challenge for ODDP. In case the mapping is available to the applications or ODDP schemes the filter tasks can potentially execute over storage drives however the data portion should be individually filterable. One exception is RAID-1, where, the data is replicated across 2 storage drives and there is no logical mapping done by RAID controller. The filter task can execute on any of the storage drives. Similarly, an ODDP scheme where the filter task executes over the RAID controller can get access to the data to be processed without accessing the logical remapping by RAID controller.

## 12 Potential Applications of ODDP

The ability of processing on storage drives may be utilized for other purposes than just filtering data for applications. In the following, we present some potential usages of such storage drives.

**Distributed RAID Controller:** The functionality of RAID can potentially be implemented by processing capable storage drives in a distributed manner. The requirement of an explicit RAID controller can be avoided as the algorithm can be implemented on drives themselves.

**Distributed Storage Controller:** On similar lines, a distributed RAID controller, the functionalities of a storage manager like a snapshot, data deduplication can also be done by a bunch of processing capable SSDs.

**Predictive Data Fetching:** Processing capability on storage drives can potentially be utilized for predicting the future data accesses. Such predictions can lead to data prefetching which can further reduce the data access times.

**The server on a Drive:** Processing capable storage drives can also implement the server functionalities. For example, a storage drive with a query processing engine can potentially act as the database server. Similarly, file servers, web servers can also be implemented on storage drives themselves.

**Self-integrating Storage:** A storage drive can discover a data storage framework and integrate itself into the system. This is similar to a scenario where a laptop discovers a wi-fi network and gets connected to the network by itself.

## 10.4 Areas Impacted by ODDP

**Newer Resource allocation schemes on the cloud:** ODDP will make resource allocations more fluid then achieved by current resource allocation mechanisms, also known as VM placement schemes [37] [38]. The current resource allocation mechanisms in the cloud do not have the option of reducing one resources allocation by increasing others. The resource requirements of VMs are rigid and are required to be allocated in requested amounts. For example, if a VM requires a certain amount of CPU, memory, network bandwidth, and storage I/O capacity, it is not possible to reduce network bandwidth or storage I/O requirement by increasing CPU and memory allocation.

ODDP changes the allocation requirements and requires new mechanism. The network bandwidth and storage I/O requirements can be reduced by allocating processing capacity on the storage drive. Similarly, CPU and memory requirements at compute node can also be reduced by relocating filter tasks to storage disks. Such exchanges may look trivial, however, it should be noted that the VM placement problem is a mix of bin packing and graph embedding problem both of which are NP-hard. Any leeway can significantly reduce the complexity and improve the efficiency of VM placement schemes.

The newer ODDP aware VM placement scheme needs to be designed and experimented with. ODDP opens a whole new field for VM placement which needs to be researched.

**Data Processing Continuum:** The storage drives are a shared resource among the applications. Multiple applications can concurrently access the data and execute filter task on the same storage drive making it overloaded. In such an overload scenario it becomes necessary to relocate some of the filter tasks back to the compute node. However, such relocation negates the benefits of ODDP. A better solution will be to relocate filter tasks from the storage drive but as close to data as possible. A probable candidate for such relocation is the storage server. There have been proposed on-storage data processing solutions (IBM Netezza [12], Oracle Exadata [11]), however, they are standalone and proprietary and incompatible with ODDP solutions.

What is required is a seamless data processing continuum which spans across different type of nodes present in the data center including storage drives, storage servers, network nodes, dedicated data filtering nodes and the compute nodes. Idea is that rather than strictly designated roles like computing nodes, network nodes and storage nodes any node which has processing capacity should be capable of processing data. ODDP can be seen as a step towards establishing such a data processing continuum.

**ODDP and IoT:** IoT [39] devices generate a time stamped log of events. These data logs are collected in repositories and further undergo analytical processing to identify patterns [40]. As the data bandwidth requirement of IoT devices is not significant, they are generally connected with low bandwidth interconnects to the internet. However, it has the growth in the number of IoT devices pose a serious capacity challenge to both the interconnects and the data repositories.

IoT can employ a model similar to ODDP where a part of analytics is performed on the IoT devices itself and only the results are sent to central repositories. Such a processing approach will not only reduce the network bandwidth requirement but also the size of data repositories.

## 13 Conclusions Further Research Challenges in ODDP

In this paper, we presented the designs, internal details, and architectures of various ODDP schemes discussed in the literature. Current data intensive applications and extremely fast secondary storage drives are a perfect fit for ODDP. However, there are still some research and developmental challenges for ODDP which needs to be addressed. Some of these challenges are mentioned below.

**ODDP Aware I/O interconnects:** Current I/O interconnects and protocols are tuned for very basic data block I/O operations (Read, Write). Moreover, the individual layers have been merged to enable faster I/O, for example, SATA carries block commands <reference>. ODDP schemes have to work around this limitation to enable communication between host agent and disk agent. Such communication approaches are not efficient and introduce delays in overall data access time. An ODDP aware I/O interconnect can significantly improve ODDP performance.

**ODDP Aware Application Programming Models:** A new programming model where applications are aware of the ODDP mechanisms is required. Applications are required to be designed in a way that they are capable of offloading or delegating a part of their data processing to storage drive whenever required. A modularized application with clear distinction of data processing tasks will benefit from ODDP. More work is required to develop such application development model. Such models can borrow from the distributed processing scheme, however, the requirements are less stringent.

**ODDP Aware I/O Schedulers:** Scheduling fitter tasks across and on ODDP enabled storage drives is an area of research. How to best distribute the filter tasks among a cluster of ODDP enabled storage drives, given the current resource utilization levels at compute nodes and storage drives themselves is going to be very important for effective ODDP installation. Such distributed tasks + I/O scheduling is a new area which has not been explored.

**Application Specific ODDP Research:** Certain widely used applications can restructure their functioning to exploit ODDP. For example, designing a new database query optimizer which can push operations to ODDP enabled storage drives.